# Superconducting quantum criticality and the anomalous scaling: A nonlinear relativistic equation

Yong Tao[†]

College of Economics and Management, Southwest University, Chongqing, China

**Abstract:** By using the Landau-Ginzburg-Wilson paradigm, we show that, near a quantum critical point (QCP), Cooper pairs at zero temperature would obey a nonlinear relativistic equation, where the imaginary time emerges as a novel dimension. This relativistic equation is applicable to certain superconductors at zero temperature for which the Faber-Pippard coherence length formula holds at and above the upper critical dimension. Here, we further show that the relativistic equation leads to a testable prediction in the vicinity of the QCP $T_c = 0$, with $T_c$ being the transition temperature. That is, for 2D overdoped (clean) superconducting films, when the parameter $T_c/(c_0 v_F)$ is lower than a characteristic scale, the Lorentz symmetry of relativistic equation arouses an anomalous scaling $\xi_0 \propto T_c^{-1.34}$, where $\xi_0$ denotes the zero-temperature coherence length, $v_F$ denotes the Fermi velocity, and $c_0$ denotes the Faber-Pippard coefficient. However, when the parameter $T_c/(c_0 v_F)$ is large enough, the Lorentz symmetry may be broken so that the Faber-Pippard scaling $\xi_0 \propto T_c^{-1}$ is restored.

[†] Correspondence to: taoyingyong@swu.edu.cn

# 1. Introduction

During recent years, quantum critical phenomena [1] have become an important topic in condensed matter physics [2, 3]. Distinguishing from thermal critical phenomena, quantum critical phenomena take place in the vicinity of zero temperature, where the imaginary time emerges as a novel dimension [1-5]. This leads to possibilities of exploring certain emergent symmetries (in high-energy physics) near a quantum critical point (QCP) [6-7], e.g., Lorentz symmetry [8, 9] and supersymmetry [3, 10]. Lorentz symmetry, as one of the fundamental principles of the relativity theory, has become a cornerstone of describing fundamental particles and fields, and all the experiments carried out thus far are consistent with it. Despite this, the origin of Lorentz symmetry still remains unknown. This partly causes the difficulty of reconciling quantum theory and relativity [11]. In this regard, some quantum gravity theories [11-13] have suggested that Lorentz symmetry is likely to be broken at an extra high energy scale, e.g., the Planck scale. However, the latest LHAASO observations [14] show that Lorentz symmetry is still maintained when approaching the Planck energy scale. This suggests that Lorentz symmetry can be regarded as a fundamental symmetry for generating a space-time. Therefore, in the vicinity of the QCP, with the emergence of the imaginary time, it is natural to conjecture that the corresponding Lorentz symmetry will emerge as well. In this paper, we further explore if an exact relativistic equation in the formalism of imaginary time would emerge in the vicinity of the QCP.

By examining 2D overdoped side of $\text{La}_{2-x}\text{Sr}_x\text{CuO}_4$ (LSCO) films, Bozovic *et al.* recently reported a two-class scaling relationship between the zero-temperature superfluid phase stiffness[1] $\rho_0$ and the transition temperature $T_c$ as below [15-17]:

$$\begin{cases} T_c = \alpha\, \rho_0 + T_0, & T_c \geq T_M \\ T_c = \gamma\, \sqrt{\rho_0}, & T_c \leq T_Q \end{cases}, \qquad (1)$$

where, $T_Q \approx 15\ \text{K}$, $T_M \approx 12\ \text{K}$, $T_0 = (7.0 \pm 0.1)\ \text{K}$, $\alpha = 0.37 \pm 0.02$, and $\gamma = (4.2 \pm 0.5)\ \text{K}^{1/2}$.

By extending Gor'kov's non-relativistic Ginzburg-Landau (GL) equation at zero

---

[1] Here, $\rho_0$ denotes the zero-temperature superfluid phase stiffness in homogenous materials [15].

temperature ($T = 0$) to the relativistic free energy (2) in the formalism of imaginary time [18-21]:

$$\mathcal{L}_q(T = 0) = |\partial_\tau \phi(\boldsymbol{q},\tau)|^2 + |\boldsymbol{\nabla}\phi(\boldsymbol{q},\tau)|^2 - \frac{24\pi^2 m_e}{7\,\zeta(3)\,\varepsilon_F} T_c^2 |\phi(\boldsymbol{q},\tau)|^2 +$$
$$\frac{12\pi^2 m_e}{7\,\zeta(3)\,\varepsilon_F} \frac{T_c^2}{\rho_0} |\phi(\boldsymbol{q},\tau)|^4, \tag{2}$$

we have shown that the experimental result (1) can be reproduced [20, 21], where the imaginary time $\tau \in [0, 1/T]$ with $T$ being the temperature, $\boldsymbol{q}$ denote the spatial coordinates, $\phi = \phi(\boldsymbol{q}, \tau)$ denotes the order parameter, $\varepsilon_F$ denotes the Fermi energy, $\zeta(3)$ denotes the Riemann zeta function, and $m_e$ denotes the mass of an electron. In this paper, we use the natural units $\hbar = c = k_B = 1$, where $\hbar$ denotes the reduced Planck constant, $k_B$ denotes the Boltzmann constant, and $c$ denotes the light speed.

In particular, by using equation (2) we further predicted that [18], for the 3D overdoped cuprate, the two-class scaling relationship (1) would be replaced by a linear scaling $T_c \propto \rho_0$, i.e., the Homes scaling. More recently, this theoretical prediction has been supported by the latest experimental data [22]. Although the validity of equation (2) has been supported by existing experiments [22, 23], we observe two possible flaws in it. First, Gor'kov [24, 25] derived the GL equation by using an $s$-wave paring BCS theory. This means that equation (2) seems to be not applicable to the $d$-wave superconductors, e.g., the LSCO cuprate. Second, Gor'kov [24, 25] derived the GL equation by using the approximation condition

$$|1 - T/T_c| \ll 1. \tag{3}$$

It can be seen that $T = 0$ breaks the inequality (3). This implies that Gor'kov's GL equation cannot be applied to the case of $T = 0$. Due to the two flaws above, the validity of equation (2) seems to be questioned. However, in this paper, we show that equation (2) can be generally derived by using the Landau-Ginzburg-Wilson paradigm so long as the Faber-Pippard coherence length formula [26] holds. By this generalized derivation, we argue that, apart from a coefficient, the relativistic equation (2) has enough universality for certain superconductors at zero temperature.

The remainder of the paper is organized as follows. In section 2, by using the Landau-Ginzburg-Wilson paradigm, we show that, near a QCP, Cooper pairs at zero temperature would obey a nonlinear relativistic equation, which is applicable to certain superconductors at zero temperature for which the Faber-Pippard coherence length formula holds at and above the upper critical dimension. In section 3, we show that the two-class scaling relationship (1) can be accounted for by this relativistic equation. In section 4, we show that the Lorentz symmetry of relativistic equation may arouse an anomalous scaling between the zero-temperature coherence length and the transition temperature. In section 5, we show that, when the Lorentz symmetry of relativistic equation is broken, the Faber-Pippard scaling is restored. In section 6, we explore the possible extension of relativistic equation for high $T_c$. Section 7 concludes the paper.

## 2. Relativistic equation for Cooper pairs at zero temperature

To derive equation (2), let us apply the Landau-Ginzburg-Wilson paradigm to the case of $T = 0$. First, we use $\rho = |\phi|^2$ to denote the zero-temperature superfluid phase stiffness so that $\phi$ is interpreted as an order parameter to describe the quantum phase transition. For example, by the experimental result (1), it can be concluded that $\rho = 0$ at $T_c = 0$. This means that the order parameter $\phi$ as a function of $T_c$ is equal to zero at $T_c = 0$; that is, $\phi(T_c = 0) = 0$ indicates the QCP [16, 17]. Second, by the Landau-Ginzburg-Wilson paradigm, we further assume that the free energy $\mathcal{F}[\phi]$ of the superconductor at zero temperature depends smoothly on the order parameter $\phi = \phi(T_c)$. Therefore, we can Taylor expand $\mathcal{F}[\phi]$ around $\phi = \phi(T_c = 0) = 0$ as the series of $|\phi|^2$:

$$\mathcal{F}[\phi] = \mathcal{F}[0] + \lambda_2(T_c)|\phi|^2 + \lambda_4(T_c)|\phi|^4 + O(|\phi|^6), \tag{4}$$

where $\lambda_2(T_c)$ and $\lambda_4(T_c)$ are two coefficients depending on the transition temperature $T_c$, $O(|\phi|^6)$ denote the higher-order terms, and $\mathcal{F}[0] = \mathcal{F}[\phi = 0]$. Here, let us firstly omit the high-order terms $O(|\phi|^6)$. In section 3, we demonstrate that the high-order terms $O(|\phi|^6)$ may be omitted when the parameter $T_c/(c_0 v_F)$ is small enough, where $v_F$ denotes the Fermi velocity and $c_0$ denotes the Faber-Pippard coefficient [refer to equation (11) later]. However, in section 6, we will explore the field equation of Cooper pairs at zero temperature when the high-order terms $O(|\phi|^6)$ cannot be omitted.

At zero temperature $T = 0$, the imaginary time $\tau$ emerges, and it in turn causes the emergence of a novel space-time[2]. Now, we assume that Lorentz symmetry is a fundamental symmetry for generating this space-time. Thus, if the order parameter $\phi$ varies in the spatial and time directions, then equation (4) can be written in the relativistic form (see page 168 in [5]):

$$\mathcal{F}[\phi(T_c)] = |\partial_\tau \phi|^2 + |\nabla \phi|^2 + \lambda_2(T_c)|\phi|^2 + \lambda_4(T_c)|\phi|^4, \tag{5}$$

where the constant term $\mathcal{F}[0]$ has been ignored simply. The Lorentz symmetry (in the formalism of imaginary time) in equation (5) is an assumption. Thus, distinguishing from the action in page 168 in [5], in equation (5) we have removed the time-derivative term $\phi^* \partial_\tau \phi$. In section 3, we show that the validity of such an assumption is justified by the experimental result (1).

To determine the coefficients $\lambda_2(T_c)$ and $\lambda_4(T_c)$, let us write down the Euler-Lagrange equation of equation (5)

$$\partial_\tau^2 \phi + \nabla^2 \phi - \lambda_2(T_c)\phi - 2\lambda_4(T_c)\phi|\phi|^2 = 0. \tag{6}$$

For the order parameter $\phi_0$ of the static and homogenous case, we should have $\partial_\tau \phi_0 = 0$ and $\nabla \phi_0 = 0$. If we order

$$\rho_0 = |\phi_0|^2, \tag{7}$$

then $\rho_0$ denotes the zero-temperature superfluid phase stiffness in the static and homogenous case. By plugging $\phi_0$ into equation (6) we have

$$\rho_0 = -\frac{\lambda_2(T_c)}{2\lambda_4(T_c)}. \tag{8}$$

---

[2] In statistical mechanics, the partition function in $D$-dimensional space is written as $Z_0(T) = \int \mathcal{D}\phi^*(\boldsymbol{q}) \int \mathcal{D}\phi(\boldsymbol{q}) \, e^{-\frac{1}{T}\int d^D q \, \mathcal{F}[\phi(\boldsymbol{q})]}$, see page 288 in [27]. If one introduces the variable $\tau \in [0, 1/T]$, then this partition function can be rewritten as $Z_0(T) = \int \mathcal{D}\phi^*(\boldsymbol{q},\tau) \int \mathcal{D}\phi(\boldsymbol{q},\tau) \, e^{-\int_0^{1/T} d\tau \int d^D q \, \mathcal{F}[\phi(\boldsymbol{q},\tau)]}$, see page 288 in [27]. By the Wick rotation $\tau = it$, the partition function becomes the path integral formula in quantum field theory. Therefore, the variable $\tau$ is interpreted as the imaginary time, see page 289 in [27]. When $T = 0$, quantum fluctuations in imaginary time need to be taken into account, and thus the imaginary time $\tau \in [0, \infty]$ effectively plays the role of an additional dimension of the system (see page 166 in [5]). That is, $(\boldsymbol{q}, \tau)$ describe a novel space-time. In this regard, Hartle and Hawking [28] have used the space-time coordinates $(\boldsymbol{q}, \tau)$ to study the wave function of the universe.

Substituting equation (8) into equation (6) yields

$$\partial_\tau^2 \phi + \nabla^2 \phi + \lambda_2(T_c)\phi\left[\frac{|\phi|^2}{\rho_0} - 1\right] = 0. \tag{9}$$

For the static case, by using equation (9) it is easy to calculate the zero-temperature coherence length [20]:

$$\xi_0 = \left(-\lambda_2(T_c)\right)^{-\frac{1}{2}}. \tag{10}$$

For clean materials, Faber and Pippard [26] have proposed that the zero-temperature coherence length can be written as

$$\xi_0 = c_0 \frac{v_F}{T_c}, \tag{11}$$

where $c_0$ is a coefficient determined by the experimental measure. Here, we call $c_0$ the Faber-Pippard coefficient, and its value is different for various materials. In contrast with the zero-temperature coherence length formula in the BCS theory [29], the Faber-Pippard formula (11) has a wider range of applicability. As we shall discuss below, the *s*-wave and *d*-wave cases can be regarded as two special examples of equation (11).

For the *s*-wave paring BCS theory, it has been known that [29] $\xi_0 = v_F/(\pi\Delta_0)$ with $\Delta_0 \approx 1.76\, T_c$, where $\Delta_0$ denotes the zero-temperature energy gap. Substituting both of them into equation (11) gives the well-known BCS result $c_0 \approx 1/(1.76\pi) \approx 0.18$ [29].

For the weak coupling model of *d*-wave superconductors, it has been theoretically shown that the zero-temperature coherence length can be written as [30]

$$\xi_0 = \frac{v_F}{\sqrt{12}\gamma_F^d \Delta_0}, \tag{12}$$

with

$$\gamma_F^d = \sqrt{\frac{\sum_k \gamma_k^2 \delta(\xi_k)}{\sum_k \delta(\xi_k)}}, \tag{13}$$

where $\xi_k$ denotes the normal-state dispersion, $\delta(\xi_k)$ denotes the Dirac function, and $\gamma_k$ denotes the factor that controls the symmetry of the BCS wave function [30].

The *s*-wave superconductors have the symmetry of $\gamma_k = 1$ [30], so, by equation (13), one has[3] $\gamma_F^d = 1$. However, for the *d*-wave case, it has been theoretically found

---

[3] If one orders $\Delta_F = \gamma_F^d \Delta_0$, then $\Delta_F$ can be interpreted as the effective gap [30]; therefore, equation (12) can be written in the form $\xi_0 = v_F/(\sqrt{12}\Delta_F)$. For the *s*-wave case ($\gamma_k = 1$), by

that [30]

$$\gamma_F^d \leq 1. \tag{14}$$

In this paper, we investigate the **d**-wave superconductors. Therefore, the inequality (14) is always taken into account. Regarding the **d**-wave superconductors, it has been experimentally found that [31, 32]

$$\frac{\Delta_0}{T_c} \approx 2.14. \tag{15}$$

Thus, by substituting equations (12) and (15) into equation (11), we obtain the Faber-Pippard coefficient for the **d**-wave superconductors as follows:

$$c_0 = \frac{1}{2.14\sqrt{12}\gamma_F^d}. \tag{16}$$

Based on the discussion above, the **s**-wave and **d**-wave cases are two special examples of equation (11).

Substituting equation (11) into equation (10) one has

$$\lambda_2(T_c) = -\frac{T_c^2}{c_0^2 v_F^2}. \tag{17}$$

By equations (8) and (17) we further obtain

$$\lambda_4(T_c) = \frac{T_c^2}{2c_0^2 v_F^2 \rho_0}. \tag{18}$$

Therefore, $\lambda_2(T_c)$ and $\lambda_4(T_c)$ are determined by equations (17) and (18). Substituting both of them into equation (5) we obtain an exact relativistic form

$$\mathcal{F}[\phi] = |\partial_\tau \phi|^2 + |\nabla \phi|^2 - \frac{T_c^2}{c_0^2 v_F^2}|\phi|^2 + \frac{T_c^2}{2c_0^2 v_F^2 \rho_0}|\phi|^4. \tag{19}$$

Equation (19) is one of two main results of this paper.

Now, we show that equation (2) is a special case of equation (19). To do so, we observe that the Fermi velocity $v_F$ is related to the electron effective mass $m_e^*$ and the Fermi energy $\varepsilon_F$. Here, the effective mass $m_e^*$ is calculated at the Fermi energy; therefore, we have the dispersion relationship of interacting electrons[4] as follows:

---

equation (13) one has $\gamma_F^d = 1$ so that $\Delta_F = \Delta_0$. Then, equation (12) agrees with the standard BCS result $\xi_0 = v_F/(\pi \Delta_0)$ except a constant factor $\sqrt{12}/\pi \approx 1.1$.

[4] For interacting electron systems, the definition of the electron effective mass $m_e^*$ at any chosen point in $k$-space for a general spherically symmetric dispersion relationship is given by $1/m_e^* = \partial^2 \varepsilon_k / \partial k^2$, where $k$ denotes the momentum and $\varepsilon_k$ denotes the energy. Thus, if the effective mass $m_e^*$ is calculated at the Fermi energy in $k$-space, then the Fermi velocity is expressed as $v_F = \partial \varepsilon_k / \partial k|_{k=k_F} = k_F/m_e^*$, where $k_F$ denotes the Fermi momentum. This means that the Fermi

$$\varepsilon_F = \frac{1}{2} m_e^* v_F^2. \tag{20}$$

By ordering $c_0 = \sqrt{\frac{7\zeta(3)}{48\pi^2}}$ and using equation (20), equation (19) becomes equation (2). However, distinguishing from equation (2), equation (19) depends on neither the BCS theory nor the approximation condition (3). By our previous derivation, the validity of equation (19) is essentially based on three assumptions:

(a). *The order parameter $\phi$ is equal to zero at $T_c = 0$.*

(b). *The Faber-Pippard coherence length formula (11) holds.*

(c). *Lorentz symmetry holds.*

Equation (19) is derived by using equation (11), which holds for clean superconductors. To take the dirty superconductors into account, in equation (10) we can simply replace $\xi_0$ by the Pippard coherence length $\xi_{dirty}$ for dirty case [33], where

$$\frac{1}{\xi_{dirty}} = \frac{1}{\xi_0} + \frac{1}{l}, \tag{21}$$

with $l$ being the mean free path. However, in this paper, we only consider the case of clean superconductors.

## 3. Reproducing the two-class scaling relationship (1)

Next, we show that the experimental result (1) can be reproduced by using equation (19). To this end, let us write down the quantum partition function of equation (5) [i.e., (19)] as:

$$Z = \int [\mathcal{D}\phi^*]_\Lambda \int [\mathcal{D}\phi]_\Lambda \, e^{-\int d\tau \int d^D q \, \mathcal{F}[\phi]}, \tag{22}$$

where $\Lambda = 1/a$ denotes the momentum cut-off and $a$ denotes the lattice spacing.

As $T_c/(c_0 v_F) \to 0$, by equation (19) quantum fluctuations will be amplified to break the mean-field approximation. Thus, by applying the renormalization group approach into equation (22), when $T_c/(c_0 v_F) \to 0$, as with the derivation in [18-21], $\lambda_4(T_c)$ in equation (5) yields a fixed point that depends on the momentum cut-off $\Lambda$:

---

energy can be written as $\varepsilon_F = (1/2) m_e^* v_F^2$; that is, equation (20).

$$\lambda_4(T_c) = (4 - D - z)\, \Lambda^{4-D-z}\, \frac{(2\pi)^D \Gamma\left(\frac{D}{2}\right)}{5(\pi)^{\frac{D}{2}}}, \tag{23}$$

where $z = 1$ denotes the dynamical critical exponent.

For $D = 2$, by using equation (18), equation (23) can be written as a parabolic scaling formula:

$$T_c = \gamma(2) \sqrt{\rho_0}, \tag{24}$$

where

$$\gamma(2) = c_0 v_F \sqrt{\frac{8\pi}{5a}}. \tag{25}$$

Here, equations (24) and (25) imply $|\phi| \propto \sqrt{\rho_0} = T_c/\gamma(2) \propto T_c/(c_0 v_F)$. This means that the high-order terms $O(|\phi|^6)$ in equation (4) may be omitted when the parameter $T_c/(c_0 v_F)$ is small enough.

Equation (24) reproduces the low $T_c$ part in the experimental result (1). Now, we further derive the high $T_c$ part in the experimental result (1). To this end, we observe that equation (24) holds when the mean-field approximation breaks down. However, when $T_c/(c_0 v_F) \gg 0$, we expect that the mean-field approximation becomes valid [18-21]. It has been known that, when the mean-field approximation holds, $\rho_0$ and $T_c$ yield a linear function [18, 20, 34, 35]:

$$T_c = \alpha\, \rho_0 + T_0, \tag{26}$$

where $\alpha$ and $T_0$ are two parameters determined by experimental measure.

To identify the range of applicability of the mean-field approximation, a quantum Ginzburg number has been proposed as [20]:

$$e^q(D) = \frac{\left|\int_0^\infty d\tau \int d^D\boldsymbol{q}\, G(\boldsymbol{q},\tau)\right|}{\int_0^\infty d\tau \int d^D\boldsymbol{q}\, \phi(\boldsymbol{q},\tau)^* \phi(\boldsymbol{q},\tau)}, \tag{27}$$

where $G(\boldsymbol{q}, \tau)$ denotes the correlation function that is defined by

$$G(\boldsymbol{q} - \boldsymbol{q}', \tau - \tau') = \langle [\phi(\boldsymbol{q},\tau) - \langle\phi(\boldsymbol{q},\tau)\rangle]\, [\phi(\boldsymbol{q}',\tau')^* - \langle\phi(\boldsymbol{q}',\tau')^*\rangle] \rangle. \tag{28}$$

By equation (27), the mean-field approximation holds if and only if

$$e^q(D) \ll 1. \tag{29}$$

By using the inequality (29), we have shown that [20] equation (26) holds when $T_c \geq T_0/(1 - \alpha)$, while equation (24) holds when $T_c \leq \gamma(2)^2$. Therefore, in the overdoped side of cuprate films ($D = 2$), it can be concluded that $\rho_0$ and $T_c$ yield a two-class relationship [20, 21]:

$$\begin{cases} T_c = \alpha\, \rho_0 + T_0, & T_c \geq T_M \approx \frac{T_0}{1-\alpha} \\ T_c = \gamma(2) \sqrt{\rho_0}, & T_c \leq T_Q \approx \gamma(2)^2 \end{cases}. \tag{30}$$

Next, we use two steps to show that the theoretical result (30) quantitatively agrees with the experimental result (1).

First, we estimate the theoretical value of $\gamma(2)$. By equation (25), it is determined by three parameters: $c_0$, $v_F$, and $a$. For the **d**-wave superconductors, we have known that $c_0$ is theoretically determined by equation (16). It has been experimentally observed [36] that the Fermi velocity of electrons may be universal for high-$T_c$ cuprates, and its estimated value is found to be $v_F \approx 3 \times 10^5\ m/s$. Moreover, Bozovic et al. have measured the lattice spacing of the LSCO as $a \approx 3.8 \times 10^{-10}$ m [15]. By substituting both data above and equation (16) into equation (25) we obtain

$$\gamma(2) \approx \frac{0.8}{\gamma_F^d}\, K^{\frac{1}{2}}. \tag{31}$$

It has been theoretically known that $\gamma_F^d \leq 1$ for the **d**-wave superconductors [30]. Thus, by equation (31) we have $\gamma(2) \geq 0.8\ K^{1/2}$, which is in the same order of magnitude as the experimental value $\gamma = (4.2 \pm 0.5)\ K^{1/2}$. In particular, if we take $\gamma_F^d \approx 1/5$, then $\gamma(2) \approx 4\ K^{1/2}$. However, the factor $\gamma_F^d$ is yet to be determined experimentally.

Equation (31) is a theoretical result of applying the weak coupling model of **d**-wave superconductors. However, by equation (25), we observe that, given the experimental value of the lattice spacing $a$, the theoretical value of $\gamma(2)$ is uniquely determined by the product of two parameters, $c_0 v_F$. More importantly, by equation (11), the product $c_0 v_F$ can be directly measured experimentally[5]. This means that by equation (25) one can estimate the theoretical value of $\gamma(2)$ by directly measuring the experimental value of $c_0 v_F$. In this sense, the validity of equation (19) can be directly tested by experiment, without involving other theoretical models. In this regard, we propose the experimentalists to check overdoped cuprate films in the vicinity of $T = 0$. For example, by equation (1) one can check the LSCO films at $T_c > 12$ K and $T < 1$K.

---

[5] Equation (11) can be generally written in the form $\xi_0 = (T_c/c_0 v_F)^{-\sigma}$. The logarithm of this function yields $\ln\xi_0 = \sigma\ln(c_0 v_F) - \sigma\ln T_c$. If one has measured the experimental values of $\xi_0$ for different $T_c$, then, by fitting these experimental values to this logarithm function, one can obtain the experimental value of $c_0 v_F$.

Second, we estimate the theoretical values of $T_M$ and $T_Q$. By using the experimental results $T_0 = (7.0 \pm 0.1)$ K, $\alpha = 0.37 \pm 0.02$, and $\gamma = (4.2 \pm 0.5)$ K$^{1/2}$ in equation (1), we have $T_M \approx T_0/(1-\alpha) \approx 11$ K and $T_Q \approx \gamma(2)^2 \approx 17$ K, which agree with the experimental values $T_M \approx 12$ K and $T_Q \approx 15$ K, respectively.

In particular, by equation (23), it can be seen that the upper critical dimension is denoted by $D_u = 3$ [18]. Therefore, the mean field approximation holds at and above $D = 3$. This theoretically predicts that [18], for 3D superconductors, the two-class relationship (30) will be replaced by the mean field result (26); that is, equation (30) is replaced by Homes' scaling [18]:

$$T_c \propto \frac{\rho_0}{\sigma_{dc}}, \qquad (32)$$

where $\sigma_{dc}$ denotes the d.c. conductivity.

This theoretical prediction in [18] has been supported by the latest experimental data [22], where Dordevic and Homes have checked thin films and bulk samples.

## 4. Lorentz symmetry and the anomalous scaling

Differing from the standard model in high energy physics, our renormalization group analyses in the section 3 imply that equation (19) has been applied to the vicinity of the cut-off $\Lambda = 1/a$, where the lattice spacing $a$ defines the range of applicability[6] of quantum field model (22). Because the theoretical result (30) quantitatively agrees with the experimental result (1), this suggests that the Lorentz symmetry of equation (19) is maintained when approaching the cut-off. This coincides with the experimental observation in high energy physics, where Lorentz symmetry is still maintained when approaching the Planck energy scale [14], which is regarded as the cut-off of the real space-time.

---

[6] This is based on the point of view from an effective field theory. Regarding this, Peskin and Schroeder have stated that (see page 402 in [37]) "any quantum field theory is defined fundamentally with a cut-off $\Lambda$ that has some physical significance. *In statistical mechanical applications, this momentum scale is the inverse atomic spacing.* In QED and other quantum field theories appropriate to elementary particle physics, the cut-off would have to be associated with some fundamental graininess of space-time." Furthermore, Weinberg also clarifies that [19] "*In solid-state physics, there really is a cut-off, the lattice spacing, which of course one must take seriously in dealing with phenomena at similar length scales.*"

However, in the real space-time, Lorentz symmetry is explicitly broken in the low-energy phases of matter. Based on this observation, we explore possible physical effects of Lorentz symmetry breaking in equation (19) in the "low-energy" phase, where the "low-energy" is defined as the low-energy (or non-relativistic) correspondence of equation (19) under a Wick rotation of imaginary time. To this end, we first check physical effects of equation (19) when the Lorentz symmetry is maintained. It can be seen that, apart from a constant, equation (19) is same as equation (2). Thus, by applying the renormalization group procedure in [21] into equation (19), as $T_c/(c_0 v_F) \to 0$, the zero-temperature coherence length $\xi_0$ and the transition temperature $T_c$ should yield a scaling relationship [21]:

$$\xi_0 \propto \left[(-\lambda_2(T_c))^{-\frac{1}{2}}\right]^\sigma = \left(\frac{T_c}{c_0 v_F}\right)^{-\sigma} \tag{33}$$

with $\sigma = 2\nu$, where $\nu$ denotes Wilson's thermal critical exponent for coherence length [38]. The presence of the factor $2$ is due to that the coefficient in front of $|\phi|^2$ in equation (19) is quadratic in $T_c$, rather than being linear in $T_c$ [21]. In existing literature, the thermal critical exponent $\nu$ has been calculated to be around 0.67 (up to the five-loop corrections) by using different methods [39-44]. In particular, the experimental measure indicates $\nu \approx 0.67$ as well [44]. Therefore, for $D = 2$, we conclude $\sigma = 2\nu \approx 1.34$. This result is derived by imposing the Lorentz symmetry on equation (19) (so that the dynamical critical exponent $z = 1$) [21]. From this sense, $\sigma \approx 1.34$ can be regarded as an indication of identifying the emergence of Lorentz symmetry for 2D superconductors. However, since the upper critical dimension is denoted by $D_u = 3$ for $z = 1$, we should expect that, for 3D superconductors, the mean-field result (11), i.e., $\xi_0 = [T_c/(c_0 v_F)]^{-1}$, always holds so that $\sigma = 1$. Henceforth, to identify the emergence of $\sigma \approx 1.34$, we always consider 2D superconductors.

## 5. Wick rotation and the non-relativistic approximation

To explore physical effects of the Lorentz symmetry breaking, we determine the "low-energy" limit of equation (19). In this paper, we define such a "low-energy" limit as the low-energy (or non-relativistic) correspondence of equation (19) under a Wick rotation of imaginary time. Thus, let us first perform the Wick rotation $\tau = it$ for equation (5) [or (19)] to obtain

$$\mathcal{L} = |\partial_t \phi|^2 - |\nabla \phi|^2 - \lambda_2(T_c)|\phi|^2 - \lambda_4(T_c)|\phi|^4, \tag{34}$$

where $\mathcal{L} = -\mathcal{F}(T_c)$.

Because $\lambda_2(T_c) < 0$ and $\lambda_4(T_c) > 0$, equation (34) is formally same as the Higgs field model (see page 227 in [27]). To find the mass term, we need to consider the spontaneous symmetry breaking in equation (34). To this end, let us first replace $\partial_t$ and $\nabla$ in equation (34) by $\partial_t \to \partial_t + ieA_0$ and $\nabla \to \nabla + ieA$, respectively, where $(A_0, A)$ denote gauge fields and $e$ denotes the charge. Thus, equation (34) can be written as

$$\mathcal{L} = |(\partial_t + ieA_0)\phi|^2 - |(\nabla + ieA)\phi|^2 - \lambda_2(T_c)|\phi|^2 - \lambda_4(T_c)|\phi|^4. \tag{35}$$

where, for simplicity, we have omitted the field strength term.

If we define the covariant derivative $\partial_\mu = (\partial_t, \nabla)$ and the covariant tensor $A_\mu = (A_0, A)$, equation (35) can be rewritten in the form:

$$\mathcal{L} = (\partial_\mu - ieA_\mu)\phi^*(\partial^\mu + ieA^\mu)\phi - \lambda_2(T_c)|\phi|^2 - \lambda_4(T_c)|\phi|^4, \tag{36}$$

where we have used the Minkowski metric

$$(g_{\mu\nu}) = (g^{\mu\nu}) = \begin{pmatrix} 1 & 0 & 0 & 0 \\ 0 & -1 & 0 & 0 \\ 0 & 0 & -1 & 0 \\ 0 & 0 & 0 & -1 \end{pmatrix} \tag{37}$$

and $\mu, \nu = 0,1,2,3$.

By plugging

$$\phi = \frac{1}{\sqrt{2}}(h + if) + \varphi_0 \tag{38}$$

into equation (36) we have

$$\mathcal{L} = \tfrac{1}{2}\partial_\mu h \partial^\mu h + \tfrac{1}{2}\partial_\mu f \partial^\mu f + \lambda_2(T_c)h^2 - \tfrac{e^2 \lambda_2(T_c)}{2\lambda_4(T_c)} A_\mu A^\mu -$$

$$\sqrt{-\lambda_2(T_c)\lambda_4(T_c)}\, h(h^2+f^2) - \tfrac{\lambda_4(T_c)}{4}(h^2+f^2)^2 + e\sqrt{\tfrac{-\lambda_2(T_c)}{\lambda_4(T_c)}}\, \partial^\mu f A_\mu +$$

$$e(h\partial_\mu f - \partial_\mu h f)A^\mu + e^2 \sqrt{\tfrac{-\lambda_2(T_c)}{\lambda_4(T_c)}}\, h A_\mu A^\mu + \tfrac{e^2}{2}(h^2+f^2) A_\mu A^\mu, \qquad (39)$$

where $h$ and $f$ are two real scalar fields, and

$$\varphi_0 = \sqrt{\tfrac{-\lambda_2(T_c)}{2\lambda_4(T_c)}}. \qquad (40)$$

By equation (39), it is easy to observe that the mass of the scalar field $h$ is denoted by

$$m_h = \sqrt{-2\lambda_2(T_c)}, \qquad (41)$$

and the mass of gauge field $A_\mu$ is denoted by

$$m_{A_\mu} = e\sqrt{-\tfrac{\lambda_2(T_c)}{\lambda_4(T_c)}}. \qquad (42)$$

Here, $f$ is the massless Goldstone boson, and it can be eliminated by using the gauge transformations $\phi' \to exp(i\gamma)\phi$ and $A'_\mu \to A_\mu - \tfrac{1}{e}\partial_\mu \gamma$. To see this, let us consider the infinitesimal transformation[7] $\phi' = exp(i\gamma)\phi \approx (1+i\gamma)\left[\tfrac{1}{\sqrt{2}}(h+if) + \varphi_0\right]$, which means the following transformations:

$$\begin{cases} h' = h - \gamma f \\ f' = f + \gamma h + \sqrt{2}\gamma \varphi_0 \end{cases}, \qquad (43)$$

where we have used equation (38) and $\phi' = \tfrac{1}{\sqrt{2}}(h'+if') + \varphi_0$.

Now, in equation (43) we choose the unitary gauge $\gamma$ to guarantee $f' = 0$; that is, $\gamma = -\tfrac{f}{h+\sqrt{2}\varphi_0}$. Thus, by using the transformation (43) under the unitary gauge, equation (39) becomes

$$\mathcal{L} = \tfrac{1}{2}\partial_\mu h \partial^\mu h + \lambda_2(T_c)h^2 - \tfrac{e^2\lambda_2(T_c)}{2\lambda_4(T_c)} A_\mu A^\mu - \sqrt{-\lambda_2(T_c)\lambda_4(T_c)}\, h^3 - \tfrac{\lambda_4(T_c)}{4} h^4 +$$

$$+e^2 \sqrt{\tfrac{-\lambda_2(T_c)}{\lambda_4(T_c)}}\, h A_\mu A^\mu + \tfrac{e^2}{2} h^2 A_\mu A^\mu, \qquad (44)$$

where, for brevity, we again use the symbol $h$ to replace $h'$.

Obviously, the Goldstone boson $f$ has vanished in equation (44). Since the mass term of the scalar field $h$ has appeared in equation (44), we can perform the non-relativistic approximation for equation (44). To this end, we use the method in existing literature [45-49] to order

$$h = \frac{1}{\sqrt{2}}(\phi + \phi^*) - \sqrt{2}\varphi_0, \tag{45}$$

where $\phi = \frac{1}{\sqrt{m_h}} exp(-im_h t)\psi$.

---

[7] The infinitesimal transformation means $|\gamma| \ll 1$ so that $exp(i\gamma) \approx (1+i\gamma)$.

Substituting equation (45) into equation (44) and by neglecting all rapidly oscillating terms [45-49] one has

$$\mathcal{L} \approx i\psi^*\partial_t\psi + \frac{1}{2m_h}\partial_t\psi^*\partial_t\psi - \frac{1}{2m_h}\boldsymbol{\nabla}\psi^*\boldsymbol{\nabla}\psi + \cdots. \tag{46}$$

From equation (46), when $|i\psi^*\partial_t\psi| \gg \left|\frac{1}{2m_h}\partial_t\psi^*\partial_t\psi\right|$, we obtain the non-relativistic approximation (see 190 in [27]). This inequality means that the low-energy limit condition of equation (44) can be denoted by

$$m_h \gg \frac{1}{2}\left|\frac{\partial_t\psi^*}{\psi^*}\right|. \tag{47}$$

By using $\phi = \frac{1}{\sqrt{m_h}}exp(-im_h t)\psi$, equation (34) can be rewritten as

$$\mathcal{L} = 2i\psi^*\partial_t\psi + \frac{1}{m_h}\partial_t\psi^*\partial_t\psi - \frac{1}{m_h}\boldsymbol{\nabla}\psi^*\boldsymbol{\nabla}\psi + \cdots. \tag{48}$$

By using the low-energy limit condition (47), equation (48) is reduced to

$$\mathcal{L} \approx 2i\psi^*\partial_t\psi - \frac{1}{m_h}\boldsymbol{\nabla}\psi^*\boldsymbol{\nabla}\psi + \cdots, \tag{49}$$

which is the low-energy limit of equation (34).

Thus, by imposing the Wick rotation $\tau = it$ on equation (49) we obtain the "low-energy" limit of equation (5) [or (19)] as below:

$$\mathcal{F}(T_c) \approx 2\psi^*\partial_\tau\psi + \frac{1}{\sqrt{-2\lambda_2(T_c)}}\boldsymbol{\nabla}\psi^*\boldsymbol{\nabla}\psi + \cdots. \tag{50}$$

For equation (50), by reference [5, 50, 51] it can be concluded that the dynamical critical exponent $z = 2$, indicating that the upper critical dimension is denoted by $D_u = 2$. This means that, in the "low-energy" limit, the mean-field approximation holds for 2D systems to guarantee $\sigma = 1$. In fact, by applying the mean-field approximation into equation (50) it is easy to calculate the zero-temperature coherence length [20]:

$$\xi_0 \propto \left(-\lambda_2(T_c)\right)^{-\frac{1}{2}} \propto T_c^{-1}, \tag{51}$$

which suggests $\sigma = 1$, indicating the Faber-Pippard scaling (11).

By using equations (17) and (41), the low-energy limit condition (47) can be rewritten as

$$\frac{\sqrt{2}T_c}{c_0 v_F} \gg \frac{1}{2}\left|\frac{\partial_t \psi^*}{\psi^*}\right|. \tag{52}$$

Inequality (52) implies that the mean field result $\sigma = 1$ holds when $T_c/(c_0 v_F)$ is large enough. However, when $T_c/(c_0 v_F)$ is less than a characteristic scale so that the inequality (52) breaks down, the Lorentz symmetry will be restored to maintain $\sigma \approx 1.34$. Therefore, for $D = 2$ we expect that $\xi_0$ and $T_c$ obey a two-class form as well. By the two-class scaling relationship (30), it can be inferred that the two-class form yields

$$\begin{cases} \xi_0 = \left(\frac{T_c}{c_0 v_F}\right)^{-1}, & T_c \geq T_M \approx \frac{T_0}{1-\alpha} \\ \xi_0 \propto \left(\frac{T_c}{c_0 v_F}\right)^{-1.34}, & T_c \leq T_Q \approx \gamma(2)^2 \end{cases}. \tag{53}$$

By inequality (52), we observe that, as with $T_c$, the product of parameters, $c_0 v_F$, should play an important role in determining the crossover point in the two-class scaling relationship (53). For example, if $c_0 v_F$ is large enough so that $T_c/(c_0 v_F)$ is sufficiently small for a given high $T_c$, then one may observe the emergence of $\sigma \approx 1.34$ at this high $T_c$ level. Conversely, if $c_0 v_F$ is sufficiently small so that $T_c/(c_0 v_F)$ for a given $T_c$ satisfies the inequality (52), then one has to seek the emergence of $\sigma \approx 1.34$ at much lower $T_c$.

## 6. The extension of equation (19) for high $T_c$

Finally, we explore the possible extension of equations (9) or (19) when $T_c$ is sufficiently larger than 0 so that the high-order terms $O(|\phi|^6)$ in Taylor series (4) cannot be omitted. As such, equation (5) can be generally written as

$$\mathcal{F}(T_c) = |\partial_\tau \phi|^2 + |\nabla \phi|^2 + \lambda_2(T_c)|\phi|^2 + \lambda_4(T_c)|\phi|^4 + \cdots + \lambda_{2n}(T_c)|\phi|^{2n}, \tag{54}$$

where $n$ is an integer.

To determine coefficients in equation (54), let us write down the Euler-Lagrange equation of equation (54):

$$\partial_\tau^2 \phi + \nabla^2 \phi - \phi[\lambda_2(T_c) + 2\lambda_4(T_c)|\phi|^2 + \cdots + n\lambda_{2n}(T_c)|\phi|^{2n-2}] = 0. \tag{55}$$

Then, we plug $\phi_0$ in equation (7) into equation (55) to obtain

$$\lambda_2(T_c)\phi_0 + 2\lambda_4(T_c)\phi_0|\phi_0|^2 + \cdots + n\lambda_{2n}(T_c)\phi_0|\phi_0|^{2n-2} = 0. \tag{56}$$

For the non-trivial solution $\phi_0 \neq 0$, equation (56) gives

$$\lambda_2(T_c) + 2\lambda_4(T_c)|\phi_0|^2 + \cdots + n\lambda_{2n}(T_c)|\phi_0|^{2n-2} = 0. \tag{57}$$

We further denote the $n-1$ solutions of equation (57) by $\eta_i$, respectively, where $i = 1,2,\ldots,n-1$. Thus, equation (57) can be written as

$$\lambda_2(T_c) + 2\lambda_4(T_c)|\phi_0|^2 + \cdots + n\lambda_{2n}(T_c)|\phi_0|^{2n-2}$$
$$= n\lambda_{2n}(T_c)[|\phi_0|^2 - \eta_1][|\phi_0|^2 - \eta_2]\cdots[|\phi_0|^2 - \eta_{n-1}] = 0. \tag{58}$$

Considering the uniqueness of the zero-temperature superfluid density (that is defined by $4m_e^*\rho_0$ [15]) for a given system, by equation (58) we assume

$$\eta_1 = \eta_2 = \cdots = \eta_{n-1} = \rho_0. \tag{59}$$

By using equation (59), equation (58) can be rewritten as

$$\lambda_2(T_c) + 2\lambda_4(T_c)|\phi_0|^2 + \cdots + n\lambda_{2n}(T_c)|\phi_0|^{2n-2} = n\lambda_{2n}(T_c)[|\phi_0|^2 - \rho_0]^{n-1}. \tag{60}$$

By using Newton's binomial formula, equation (60) can be written as

$$\lambda_2(T_c) + 2\lambda_4(T_c)|\phi_0|^2 + \cdots + n\lambda_{2n}(T_c)|\phi_0|^{2n-2} =$$
$$[\sum_{k=0}^{n-1} C_{n-1}^k |\phi_0|^{2k} n\lambda_{2n}(T_c)(-\rho_0)^{n-k-1}]. \tag{61}$$

Comparing both sides on equation (61), one has

$$\lambda_2(T_c) = n\lambda_{2n}(T_c)(-\rho_0)^{n-1}. \tag{62}$$

Because $\lambda_j(T_c)$ for $j = 2,4,\ldots,2n$ is not a function about $\tau$ and $\boldsymbol{q}$, we can simply replace $\phi_0$ by $\phi$ in equation (60) to rewrite equation (55) as

$$\partial_\tau^2 \phi + \nabla^2 \phi - n\lambda_{2n}(T_c)\phi[|\phi|^2 - \rho_0]^{n-1} = 0. \tag{63}$$

Thus, substituting equation (62) into equation (63) yields

$$\partial_\tau^2 \phi + \nabla^2 \phi + \frac{T_c^2}{c_0^2 v_F^2}\phi\left[1 - \frac{|\phi|^2}{\rho_0}\right]^{n-1} = 0, \tag{64}$$

where $\lambda_2(T_c)$ has been denoted by equation (17). Equation (64) is an extension of equation (9) when $T_c$ is sufficiently larger than 0. In particular, when $n = 2$, equation (64) becomes equation (9). To explore non-trivial topological phenomena[8], it is possible to consider $n > 2$ [58]. However, for the case of $T_c \gg 0$, we propose that the value of $n$ in equation (64) should be determined by the experimental measure. By using equation (64), the free energy (19) can be extended to the general form:

$$\mathcal{F}(T_c) = |\partial_\tau \phi|^2 + |\nabla \phi|^2 + \frac{T_c^2 \rho_0}{n c_0^2 v_F^2}\left[1 - \frac{|\phi|^2}{\rho_0}\right]^n, \tag{65}$$

which is the second one of two main results of this paper. Equation (65) becomes equation (19) when $n = 2$. In section 5, we have pointed out that, under the Wick rotation $\tau = it$, equation (19) is formally same as the Higgs field model. In fact, Higgs mode in superconductors has become an important topic [59-63].

---

[8] It's worth mentioning that the topological superconductivity has become a hot topic [52-57].

Finally, we observe that equation (64) is derived in a Euclidean space. It is possible to extend equation (64) in a fractal space. In this regard, certain scholars have studied fractal superconductivity [64-68] and impurities [69].

## 7. Conclusion

In conclusion, by using the Landau-Ginzburg-Wilson paradigm, we have theoretically shown that, near a QCP, Cooper pairs at zero temperature would obey a nonlinear relativistic equation, where the imaginary time emerges as a novel dimension. We argue that this relativistic equation is applicable to certain superconductors at zero temperature for which the Faber-Pippard scaling $\xi_0 \propto T_c^{-1}$ holds at and above the upper critical dimension. However, when the dimension of superconductor is lower than the upper critical dimension, the Faber-Pippard scaling might break down. In particular, for 2D overdoped superconducting films, we predict that, when the parameter $T_c/(c_0 v_F)$ is lower than a characteristic scale, the Lorentz symmetry of relativistic equation would arouse an anomalous scaling $\xi_0 \propto T_c^{-1.34}$. This leads to possibilities for experimentally testing certain overdoped superconducting films at zero temperature. For example, if $c_0 v_F$ is large enough so that $T_c/(c_0 v_F)$ is sufficiently small for a given high $T_c$, then one may observe the emergence of the anomalous exponent $\sigma \approx 1.34$ at this high $T_c$ level. Conversely, if $c_0 v_F$ is sufficiently small so that $T_c/(c_0 v_F)$ for a given $T_c$ satisfies the inequality (52), then one has to seek the emergence of $\sigma \approx 1.34$ at much lower $T_c$.

**References:**

[1]. J. A. Hertz, Quantum critical phenomena. *Phys. Rev. B* **14**, 1165 (1976)

[2]. I. Boettcher and I. F. Herbut, Superconducting quantum criticality in three-dimensional Luttinger semimetals. *Phys. Rev. B* **93**, 205138 (2016)

[3]. N. Zerf, C. H. Lin, and J. Maciejko, Superconducting quantum criticality of topological surface states at three loops. *Phys. Rev. B* **94**, 205106 (2016)

[4]. A. Connes and C. Rovelli, Von Neumann algebra automorphisms and time-thermodynamics relation in generally covariant quantum theories. *Classical and Quantum Gravity* **11**, 2899 (1994)

[5]. I. Herbut, *An Modern Approach to Critical Phenomena* (Cambridge University Press, 2007)

[6]. G. Chapline et al., Quantum phase transitions and the breakdown of classical general relativity. *Philosophical Magazine B*, **81**, 235-254 (2001)

[7]. R. B. Laughlin, Emergent relativity. *International Journal of Modern Physics A* **18**, 831-853 (2003)

[8]. B. Roy, V. Juricic, and I. F. Herbut, Emergent Lorentz symmetry near fermionic quantum critical points in two and three dimensions. *Journal of High Energy Physics* **2016**, 18 (2016)

[9]. V. A. Kostelecky et al., Lorentz violation in Dirac and Weyl semimetals. *Phys. Rev. Research* **4**, 023106 (2022)

[10]. S. S. Lee, TASI Lectures on Emergence of Supersymmetry, Gauge Theory and String in Condensed Matter Systems. arXiv:1009.5127

[11]. J. Collins, et al., Lorentz invariance and quantum gravity: an additional fine-tuning problem? *Phys. Rev. Lett.* **93**, 191301 (2004)

[12]. P. Horava, Quantum gravity at a Lifshitz point. *Phys. Rev. D* **79**, 084008 (2009)

[13]. A. Wang, Horava gravity at a Lifshitz point: A progress report. *International Journal of Modern Physics D* **26**, 1730014 (2017)

[14]. Z. Cao et al., Exploring Lorentz Invariance Violation from Ultrahigh-Energy $\gamma$ Rays Observed by LHAASO. *Phys. Rev. Lett.* **128**, 051102 (2022)

[15]. I. Božović, et al., Dependence of the critical temperature in overdoped copper oxides on superfluid density, *Nature* **536**, 309-311 (2016)

[16]. I. Božović, et al., The Vanishing Superfluid Density in Cuprates-and Why It matters. *Journal of Superconductivity and Novel Magnetism* **31**, 2683-2690 (2018)

[17]. I. Božović, et al., Can high-$T_c$ superconductivity in cuprates be explained by the conventional BCS theory? *Low Temperature Physics* **44**, 519-527 (2018)

[18]. Y. Tao, BCS quantum critical phenomena. *Europhysics Letters* **118,** 57007 (2017)

[19]. Y. Tao, Parabolic Scaling in Overdoped Cuprate Films. *Journal of Superconductivity and Novel Magnetism* **32**, 3773-3777 (2019)

[20]. Y. Tao, Parabolic Scaling in Overdoped Cuprate: a Statistical Field Theory Approach. *Journal of Superconductivity and Novel Magnetism* **33**, 1329-1337(2020)

[21]. Y. Tao, Relativistic Ginzburg-Landau equation: An investigation for overdoped cuprate films. *Physics Letters A* **384**, 126636 (2020)

[22]. S. V. Dordevic and C. C. Homes. Superfluid density in overdoped cuprates: Thin films versus bulk samples. *Phys. Rev. B* **105**, 214514 (2022)

[23]. C. Herrera et al., Scanning SQUID characterization of extremely overdoped $La_{2-x}Sr_xCuO_4$. *Physical Review B* **103**, 024528 (2021)

[24]. L. P. Gor'kov, Microscopic derivation of the Ginzburg-Landau equations in the theory of superconductivity. *Soviet Phys. JETP* **9**, 1364-1367 (1959)

[25]. A. A. Abrikosov, L. P. Gor'kov and I. E. Dzyaloshinskii, *Methods of Quantum Field Theory in Statistical Physics* (Prentice-Hall, Englewood Cliffs, NJ, 1963)

[26]. T. E. Faber and A. B. Pippard, The penetration depth and high-frequency resistance of superconducting aluminium. *Proc. Roy. Soc. (London) A* **231**, 53 (1955)

[27]. A. Zee, Quantum Field Theory in a Nutshell (Second Edition). (Princeton University Press, 2010)

[28]. J. B. Hartle and S. W. Hawking, Wave function of the Universe. *Phys. Rev. D* **28**, 2960 (1983)

[29]. J. Bardeen, L. N. Cooper, and J. R. Schrieffer, Theory of Superconductivity. *Phys. Rev.* **108**, 1175 (1957)

[30]. L. Benfatto et al., Coherence length in superconductors from weak to strong

coupling. *Phys. Rev. B* **66**, 054515 (2002)

[31]. H. Won and K. Maki, D-wave superconductor as a model of high Tc superconductor. *Physica B: Condensed Matter Volumes* **194-196, Part 2**, 1459-1460 (1994)

[32]. Y. Wang et al., Weak-coupling $d$-wave BCS superconductivity and unpaired electrons in overdoped $La_{2-x}Sr_xCuO_4$ single crystals. *Phys. Rev. B* **76**, 064512 (2007)

[33]. A. B. Pippard, An experimental and theoretical study of the relation between magnetic field and current in a superconductor. *Proc. Roy. Soc. (London) A* **216**, 547 (1953).

[34]. V. G. Kogan, Homes scaling and BCS, *Physical Review B* **87**, 220507(R) (2013)

[35]. V. A. Khodel, et al., Impact of electron-electron interactions on the superfluid density of dirty superconductors. *Phys. Rev. B* **99**, 184503 (2019)

[36]. X. J. Zhou, et al., Universal nodal Fermi velocity. *Nature* **423**, 398 (2003)

[37]. M. E. Peskin and D. V. Schroeder, *An Introduction to Quantum Field Theory* (Beijing World Publishing Corp) 2006

[38]. K. G. Wilson and J. B. Kogut, The renormalization group and the epsilon expansion. *Physics Reports* **12**, 75-200 (1974)

[39]. H. Kleinert, et al., Five-loop renormalization group functions of $O(n)$-symmetric $\phi^4$-theory and $\epsilon$-expansions of critical exponents up to $\epsilon^5$. *Physics Letters B* **272**, 39-44 (1991)

[40]. H. Kleinert, Strong-coupling $\phi^4$-theory in $4-\epsilon$ dimensions, and critical exponents. *Physics Letters B* **434**, 74-79 (1998)

[41]. H. Kleinert, Critical exponents without b-function. *Physics Letters B* **463**, 69-76 (1999)

[42]. H. Kleinert, Strong-coupling behavior of $\phi^4$ theories and critical exponents. *Physical Review D* **57**, 2264 (1998)

[43]. H. Kleinert and V. Schulte-Frohlinde, Critical exponents from five-loop strong-coupling $\phi^4$-theory in $4-\epsilon$ dimensions. *Journal of Physics A: Mathematical and General* **34**, 1037-1049 (2001)

[44]. R. Guida and J. Zinn-Justin, Critical exponents of the $N$-vector model. *Journal*

*of Physics A: Mathematical and General* **31**, 8103 (1998)

[45]. A. H. Guth, M. P. Hertzberg, and C. P. Weinstein, Do dark matter axions form a condensate with long-range correlation? *Phys. Rev. D* **92**, 103513 (2015)

[46]. M. H. Namjoo, A. H. Guth, and D. I. Kaiser, Relativistic corrections to nonrelativistic effective field theories. *Phys. Rev. D* **98**, 016011 (2018)

[47]. J. Eby et al., Classical nonrelativistic effective field theory and the role of gravitational interactions. *Phys. Rev. D* **99**, 123503 (2019)

[48]. L. H. Heyen and S. Floerchinger, Real scalar field, the nonrelativistic limit, and the cosmological expansion. *Phys. Rev. D* **102**, 036024 (2020)

[49]. B. Salehian et al., Beyond Schrödinger-Poisson: nonrelativistic effective field theory for scalar dark matter. *Journal of High Energy Physics* **2021**, 50 (2021)

[50]. M. P. A. Fisher et al., Boson localization and the superfluid-insulator transition. *Phys. Rev. B* **40**, 546 (1989)

[51]. S. Sachdev, *Quantum Phase Transitions* (Cambridge University Press, Cambridge, 1999).

[52]. K. H. Wong et al., Higher order topological superconductivity in magnet-superconductor hybrid systems. *npj Quantum Materials* **8**, 31 (2023)

[53]. M. Bazarnik et al., Antiferromagnetism-driven two-dimensional topological nodal-point superconductivity. *Nature Communications* **14**, 614 (2023)

[54]. S. Ono et al., Refined symmetry indicators for topological superconductors in all space groups. *Science Advances* **6**, eaaz8367 (2020)

[55]. J. Li et al., Two-dimensional chiral topological superconductivity in Shiba lattices. *Nature Communications* **7**, 12297 (2016)

[56]. B. W. Heinrich et al., Single magnetic adsorbates on s-wave superconductors. *Progress in Surface Science* **93**, 1-19 (2018)

[57]. Z. B. Yan, Higher-Order Topological Odd-Parity Superconductors. *Phys. Rev. Lett.* **123**, 177001 (2019)

[58]. B. A. Malomed and A. A. Nepomnyashchy, Kinks and solitons in the generalized Ginzburg-Landau equation. *Phys. Rev. A* **42**, 6009 (1990)

[59]. L. Schwarz et al., Classification and characterization of nonequilibrium Higgs

modes in unconventional superconductors. *Nature Communications* **11**, 287 (2020)

[60]. D. Pekker1 and C. M. Varma, Amplitude/Higgs Modes in Condensed Matter Physics. *Annual Review of Condensed Matter Physics* **6**, 269-297 (2015)

[61]. H. Krull et al., Coupling of Higgs and Leggett modes in non-equilibrium superconductors. *Nature Communications* **7**, 11921 (2016)

[62]. A. Moor et al., Amplitude Higgs Mode and Admittance in Superconductors with a Moving Condensate. *Phys. Rev. Lett.* **118**, 047001 (2017)

[63]. T. Cea et al., Nonlinear optical effects and third-harmonic generation in superconductors: Cooper pairs versus Higgs mode contribution. *Phys. Rev. B* **93**, 180507(R) (2016)

[64]. R. A. El-Nabulsi, Superconductivity and nucleation from fractal anisotropy and product-like fractal measure. *Proceedings of the Royal Society A* **477**, 20210065 (2021)

[65]. R. A. El-Nabulsi and W. Anukool, On nonlocal Ginzburg-Landau superconductivity and Abrikosov vortex. *Physica B* **644**, 414229 (2022)

[66]. R. A. El-Nabulsi and W. Anukool, Some new aspects of fractal superconductivity. *Physica B* **646**, 414331 (2022)

[67]. R. A. El-Nabulsi and W. Anukool, On fractal thermodynamics of the superconducting transition and the roles of specific heat, entropy and critical magnetic field in disordered superconductors. *Physica C* **611**, 1354302 (2023)

[68]. R. A. El-Nabulsi, Nonlocal-In-Time kinetic energy description of superconductivity. *Physica C* **577**, 1353716 (2020)

[69]. T. Jujo, Quasiclassical Theory on Third-Harmonic Generation in Conventional Superconductors with Paramagnetic Impurities. *J. Phys. Soc. Jpn.* **87**, 024704 (2018)